# Bound Pairs: Direct Evidence for Long-range Attraction between Like-Charged Colloids

B.V.R. Tata, P.S. Mohanty, and M.C. Valsakumar

Materials Science Division, Indira Gandhi Centre for Atomic Research,
Kalpakkam – 603102, Tamil Nadu, India

We report observations of stable bound pairs in very dilute deionized aqueous suspensions of highly charged polystyrene colloidal particles, with monovalent counterions, using a confocal laser scanning microscope. Through an analysis of several thousands of time series of confocal images recorded deep inside the bulk suspension, we find that the measured pair-potential, $U(r)$ has a long-range attractive component with well depths larger than the thermal energy. These observations provide a direct and unequivocal evidence for the existence of long-range attraction in $U(r)$ of like-charged colloidal particles.

Electrostatic interactions between charged macromolecules dispersed in a polar solvent have attracted sustained cross-disciplinary interest because of their fundamental role in governing the physical properties of *soft matter viz.*, colloidal suspensions, polyelectrolyte solutions and many biological systems [1]. Among the various soft matter systems, charge-stabilized colloidal suspensions (e.g., micron or submicron sized polystyrene latex and silica particles dispersed in water) have gained recognition as tremendously useful model condensed matter systems because of their structural ordering and rich phase behavior [2]. Ensembles of uniformly sized charged colloidal particles undergo phase transitions from gas to liquid, fluid to solid, and from crystal to glass [2]. Such structural orderings and phase transitions are amongst the phenomena that colloidal systems have in common with atomic systems. Hence, these systems are of considerable interest to study fundamental questions, which are also relevant for atomic systems. Understanding of this simple model system is, however, far from complete with respect to the effective inter-particle interaction between the charged colloids arising due to the basic electrostatic interactions among the constituents. Unlike the atoms, colloidal particles deep inside a suspension can be tracked with a confocal laser scanning microscope (CLSM). This experimental accessibility provides unparalleled opportunities for microscopic study of mechanisms of ordering and phase transitions [2].

The phenomena such as aggregation, fluid-crystal and body-centered-cubic (bcc) to face-centered-cubic (fcc) phase transitions, observed in bulk charged colloidal dispersions [2] have been understood using Derjaguin-Landau-Verwey-Overbeek (DLVO) theory [3]. In this theory, a pair of spherical like-charged colloidal particles immersed in an electrolyte solution repel each other at large inter-particle distances, *r,* due to screened Coulomb repulsion; they can experience strong attraction (van der Waals attraction) at very short surface-to-surface distances (~1 nm) in the presence of high salt concentration. Nevertheless it failed to explain several new and important observations [4-7]. The most striking of these observations include equilibrium phase separation (gas-liquid) between colloidal fluids of widely different densities [4], reentrant order-disorder transition [5] and stable voids coexisting with dense ordered or disordered regions (gas-solid coexistence) [6, 7]. None of these should be possible in a system with purely repulsive interactions, and hence their occurrence suggests existence of long-range attraction between like-charged colloids. The need to proceed beyond the DLVO theory has been recognized on theoretical grounds as well [8-11].

The puzzling observations in bulk suspensions mentioned above, as well as measurements of attractive interaction in the effective pair-potential $U(r)$ of confined colloidal particles [12-14], have raised a debate about its existence in the $U(r)$ of like-charged colloids under *confinement-free* conditions. For a very dilute suspension, the pair-correlation function $g(r)$ is directly related to $U(r)$ by

$$g(r) = \exp[-U(r)/k_BT] \qquad (1)$$

Hence, $U(r)$ can be obtained by measuring $g(r)$. Optical microscopy and optical tweezer techniques have been used to measure $g(r)$ by imaging the colloidal particles [12-14]. However, *only* the particles close to the microscope cover-glass could be imaged in these studies because of the limited depth of focus. The cover-glass develops negative charge on contact with water, and its close proximity to the charged colloidal particles has strong influence on $U(r)$. Thus, these studies allow determination of the effective interaction between the colloidal particles under *confinement*. It must also be mentioned that the attraction observed in the effective pair-potential using optical tweezer technique can be accounted for by a nonequilibrium hydrodynamic effect [14,15]. Thus, to date, there is no direct experimental evidence for existence of a long-range attractive term in the $U(r)$ of like-charged colloids in a bulk (*confinement-free*) suspension. In this letter we report direct evidence for long-range attraction between like-charged polystyrene spheres in a very dilute deionized bulk suspensions with *monovalent counterions* and show the existence of long-range attractive term in the measured $U(r)$.

The possibility of obtaining an attractive interaction mediated by counterions is recognized for quite some time [8]. Mean field calculations based on Poisson-Boltzmann approach are not adequate in the limit of high charge densities where counterion-counterion correlations and nonlinear effects are important. The calculations that incorporate these correlations clearly demonstrate new phenomena such as counterion condensation, charge reversal, etc [10,11]. Some of these studies demonstrate occurrence of an effective short-range attractive interaction between the macroions [10,11]. There are a few theoretical studies that suggest occurrence of a long ranged attractive interaction between like charged colloids [16,17] in the strong coupling limit $\Gamma >> 1$ ($\Gamma = \lambda_B \sqrt{z^3 \sigma / e}$, where $z$, valency of the counterion, $\sigma$, the bare charge density of colloidal particles, $e$ is the electron charge and Bjerrum length, $\lambda_B = e^2/4\pi\varepsilon_0\varepsilon k_BT$. Here, $\varepsilon_0$ is the vacuum permittivity and $\varepsilon$ is the dielectric constant of the solvent).

Observation of long-range attraction in very dilute bulk suspensions requires setting-up right experimental conditions: (a) Particles of high effective charge density, $\sigma_e$ are essential to increase the strength of attraction as well as to increase the counterion concentration in the medium, which mediate the attraction between like-charged particles [8]. (b) Elimination of wall effects and (c) avoiding sedimentation of particles due to gravity. We have selected highly charged polystyrene colloidal particles among our stock of synthesized suspensions to increase the concentration of counterions and redispersed in a density-matched fluid (50:50 $H_2O$-$D_2O$ mixture) to eliminate sedimentation due to gravity. The effect of charged wall is eliminated by employing a confocal laser scanning microscope (inverted type, Leica TCS-SP2-RS, Germany) for imaging the particles far away (> 200 μm) from charged cover slip.

We minimize the impurity ion concentration by completely deionizing the suspension using mixed-bed ion-exchange resins. We report here for the first time, the observation of bound pairs of like-charged colloidal particles separated by more than a micrometer distance and stable over several seconds. The pair-potential $U(r)$ obtained from the measured pair correlation function $g(r)$ shows an attractive minimum at an interparticle distance of more than a micrometer. These observations provide direct evidence for existence of long-range attraction in $U(r)$ of like-charged colloids. In order to establish that the above results are not experimental artifacts, we have carried out a study on a colloidal suspension with extremely low charge density particles and reproduced the standard screened coulomb repulsive interaction.

The monovalent counterion concentration in the suspension is increased by choosing highly charged polystyrene particles with strongly acidic groups on the surfaces. To examine the dependence of attractive minimum on the effective charge density $\sigma_e$, we have prepared two suspensions with particles differing in $\sigma_e$ but having same diameter $d = 0.6$ μm. The two suspensions were synthesized by emulsifier-free emulsion copolymerization of styrene and potassium styrene-sulfonate with $K_2S_2O_8$ as initiator in water at 70°C under nitrogen atmosphere and thoroughly purified using the procedure described elsewhere [7]. Samples prepared in this way are free from tethering [7]. The effective surface charge density on the particles of the two suspensions, determined using conductivity method, is 2.7 μC/cm$^2$ and 0.3 μC/cm$^2$ (~ 25 times smaller than $\sigma$). Very dilute samples S1 ($\sigma_e$ = 2.7 μC/cm$^2$) and S2 ($\sigma_e$ = 0.3 μC/cm$^2$) with a volume fraction $\phi$ of 0.0001 were prepared in a quartz container with density matching medium and purified using mixed-bed ion-exchange resins. The counterions (H$^+$) in purified samples are monovalent. We have also prepared another density matched, deionized dilute sample S3 ($\phi$ = 0.0001, d = 780 nm, $\sigma_e$ = 0.004 μC/cm$^2$) with very low surface charge density polystyrene particles (purchased from Interfacial Dynamics Corporation, Portland, OR) having weakly acidic carboxylate end groups. All the samples were shaken repeatedly and left undisturbed for a week for deionization to complete. From the conductivity measurements, the residual impurity ion concentration in our samples is estimated to be about 0.5 μM. After reaching deionization equilibrium 1 ml of samples were transferred into clean cylindrical quartz sample cells (25 mm in height, 8 mm in diameter with cover glass fixed to one end) and sealed hermetically using parafilm [7]. The sealed cells were kept at constant temperature ($T$ = 25°C) and were mounted on CLSM platform for observation.

We view the colloidal particles with a fast (7.3 frames/sec) scanning CLSM to obtain two dimensional (2D) optical slices, at a depth of 200 μm from the cover-slip to eliminate the wall effects. We record several thousands of images as time series and several such series from different locations in the sample. We obtain in-plane $g(r)$ (accounting for the finite thickness of the optical slices [18]) by determining the positions of the particles by image analysis in each frame, and then averaging over several thousands of such frames. The $U(r)$ is obtained from the measured $g(r)$ using Eq. 1. The $g(r)$ and $U(r)$ thus obtained for sample S3 are shown in Fig. 1A. Notice that the $U(r)$ is purely repulsive and fits to the Yukawa form $U(r) = A\, exp(-\kappa r)/r$. The fitted values of the Coulomb potential strength A, and the inverse Debye screening length κ are 1.2 ± 0.2 × 10$^{-25}$ J.m and 2.1 ± 0.2 μm$^{-1}$, respectively, and these values are reasonable. Since particles of sample S3 are of low charge density, the counterion concentration in the sample is much smaller (more than three orders of magnitude) than the residual salt concentration. Under these conditions DLVO theory is expected to provide an accurate description of the pair-interactions between charged colloids, and we indeed obtain the DLVO potential from our analysis of the experimental data. Further, absence of a minimum in *U(r)* at distances *r* < 2*d* implies that our measurements are free from imaging artifacts [19].

Equation 1 relates the interaction between particles at a separation *r* in three dimensions provided *g(r)* is the bulk pair correlation function. Though our measurements are on bulk suspensions, the measured *g(r)* (for example, Fig. 1A) actually corresponds to optical slices of thickness (~ 0.9 μm). In order to demonstrate that these two *g(r)*s ( bulk and slice) are the same for a dilute suspension, we have performed Monte-Carlo (MC) simulations [19] using DLVO potential with the suspension parameters same as those of sample S3. For simulating a bulk system we use periodic boundary conditions and for calculating in-plane *g(r)* we divide the simulation cell along the z-direction into several slices with thickness close to the optical slice thickness in the experiment. In order to have sufficient number of particles in a slice we fix number of particles *N* in the simulation cell to be 16000. After reaching equilibrium, we calculate the bulk *g(r)* using particle coordinates in the entire simulation cell and in-plane *g(r)* using particle coordinates in a given slice. The two *g(r)*s (bulk and slice) and the corresponding *U(r)*s are shown in Fig. 1B. Notice that both *g(r)*s as well as *U(r)*s are identical within statistical accuracy. Hence we conclude that measuring in-plane *g(r)* is sufficient for determining U(r). Thus, through our measurements as well as simulations on low charge density particles, we have established a methodology for determining *U(r)* that is free from projection errors and optical artifacts.

Using the same methodology, we have carried out CLSM studies on highly charged particles (sample S1 and S2). During these experiments, we had observed bound pairs (not touching each other) (Fig. 2) undergoing Brownian motion in the observation plane. Since we record time series of images at an interval of 137 msec (1 frame time), we could track several pairs in the field of view [21]. Some of them have stayed in the field of view up to 7 seconds and hence, they are stable up to a minimum of 7 s (~51 frames). For the sake of clarity we presented in Fig. 2 only 5 frames (not consecutive) instead of all 51 frames. We have also observed three, four and five particle clusters (Fig. 3), relatively less in number and stable over several seconds [21]. These clusters are found to change their shape and also undergo change in size due to association or dissociation of a particle from the cluster [21]. Hence these clusters are not rigid. These observations constitute direct and unambiguous evidence for the existence of long-range attraction between like-charged colloids.

Existence of such stable bound pairs implies an attractive minimum $U_m$ with magnitudes of the order of thermal energy $k_BT$ or more in the pair-interaction *U(r)* between the like-charged particles. The magnitude of $U_m$ and the interparticle distance $R_m$ at which this attractive minimum occurs, have been measured by obtaining *U(r)* using Eq. (1). The *g(r)* and U(r) for two samples S1 and S2 is shown in Fig. 4. The pair- potential *U(r)* clearly shows an attractive minimum (Fig. 4) at an interparticle distance of 1.45 μm (sample S1) and 1.48μm (samples S2), which is really long-ranged as it corresponds to about 2.4 times the diameter of the particles. Further, the well depth (Fig. 4) is found to be larger ($U_m$ ~ 1.82 $k_BT$) for particles of higher charge density (sample S1) as compared to that for particles of sample S2 ($U_m$ ~ 1.46 $k_BT$), which have lower charge density. Our measurements thus confirm the existence of a long-range attractive term in the effective pair-potential between like-charged colloids.

Measured experimental *U(r)* data was found to fit to the following empirical expression

$$U(r) = A\frac{exp(-\kappa r)}{r} - B exp(-\kappa r) \qquad (2)$$

where the constants *A*, *B* represent the strength of repulsive and attractive components of the interaction, respectively and $\kappa^{-1}$ represents the range of interaction. The fit to Eq. 2 is shown in Fig. 3. Though Eq. 2 resembles the Sogami-Ise pair-potential [8], the constants *A*, *B* and *κ* are independent here. The values of the constants *A, B* and *κ,* obtained from the fit, are higher for sample S1 ($A = 1.8 \pm 0.5 \times 10^{-21}$ J.m, *B*: $1.4 \pm 0.4 \times 10^{-15}$J, $\kappa = 6.75 \pm 0.2$ μm$^{-1}$) as compared to that for sample S2 ($A = 9.9 \pm 1.9 \times 10^{-23}$ J.m, *B*: $7.5 \pm 1.6 \times 10^{-17}$ J, $\kappa = 4.87 \pm 0.14$ μm$^{-1}$) and is due to higher charge density of the particles and the associated higher counterion concentration in the sample S1.

There is as yet no theory that can be used to readily interpret the observed results. It is widely believed that polyvalency of counterions is a must for attractive interaction between like-charged macroions in aqueous suspensions [10,11]. However, we have shown here that attraction is possible with monovalent counterions. The *Γ* values computed using the bare charge density (obtained from the measured effective charge density and the reported degree of dissociation for latex suspensions [7]) of particles are 1.29 and 0.43 for samples S1 and S2 respectively and these values are far away from the high coupling limit (*Γ* >>1), where short-range attraction is known to arise in charged systems with divalent counterions [16,17]. We note that the *imag*e charge effects [10, 22] arising due to the difference (Δε ~77.5) in the dielectric constants between the particle and the solvent can be significant for aqueous suspensions. We believe that calculations, which take into account counterion-counterion correlations along with the image charge effects, may provide explanation for the present observations.

To conclude, the direct observation of bound pairs and other stable small particle clusters using a confocal laser scanning microscope, and measurement of *U(r)* using Eq. 1 and the coordinates of millions of particles in a very dilute bulk suspension of highly charged particles allowed us to establish existence of long-range attraction in the effective pair-potential of like-charged colloidal particles. Though counterions mediate the attraction, the exact mechanism still remains as an open question. Thus, these results provide crucial experimental input for theoretical modeling of effective interactions in charged colloids and other strongly charged soft matter.

We acknowledge Junpei Yamanaka for giving us aqueous suspensions of highly charged particles. Special thanks are due to M. Muthukumar for helpful discussions. We thank, C.S. Sundar and A.K. Arora for useful suggestions.

───────────────────────────

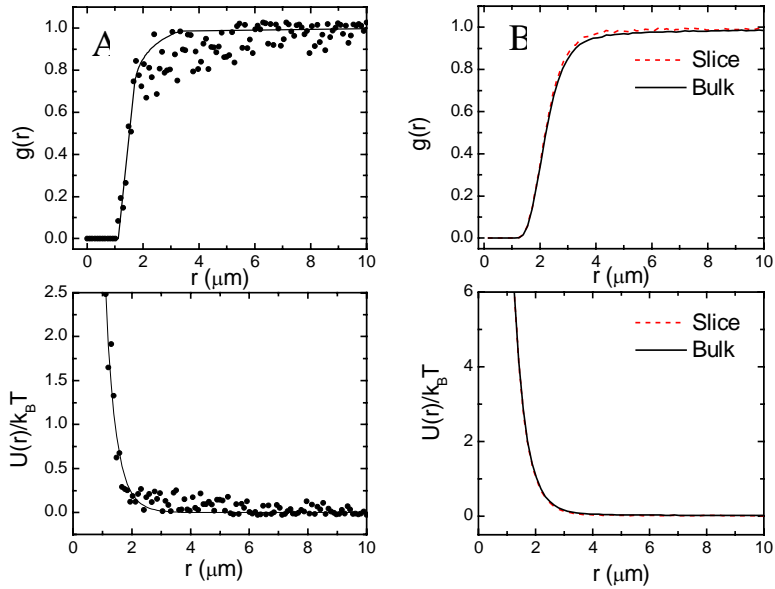

Fig. 1. (A) The measured pair-correlation function *g(r)* and the corresponding effective pair-potential *U(r)* (in units of $k_BT$) determined using Eq. (1) for colloidal particles of very low charge density (sample S3). *U(r)* is purely repulsive continuous line is the fit to the screened Coulomb Yukawa potential (see text). The line drawn through the *g(r)* data is guide to the eye. (B) *g(r)* and the corresponding *U(r)* obtained through MC simulations for a bulk suspension (continuous line) and for a slice (dotted line) of thickness ~ 0.9 μm.

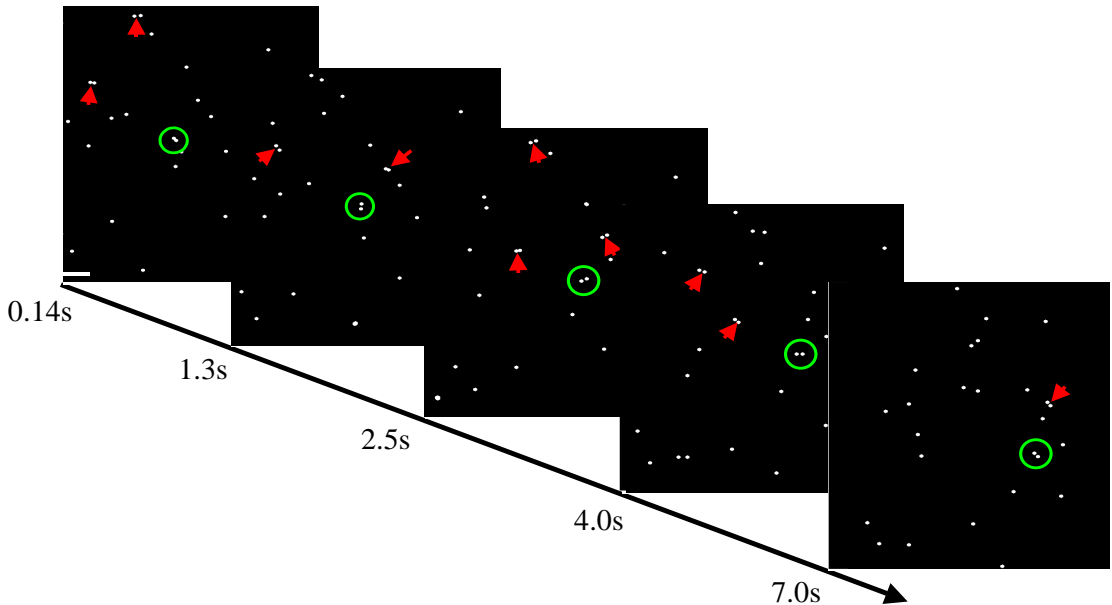

Fig. 2. Stability of bound pairs. Confocal images recorded in sample S1 as a function of time at a depth of 200 μm from the cover-slip using a 40x/0.75 objective. The arrows indicate bound pairs of polystyrene particles. A bound pair located almost at the center of the image (marked by a circle) is found to move within the field of view up to 7 s (see text) before it moved out of field of view. Other bound pairs marked by arrows are found to move out from the field of view at much shorter times. The scale bar corresponds to 5 μm.

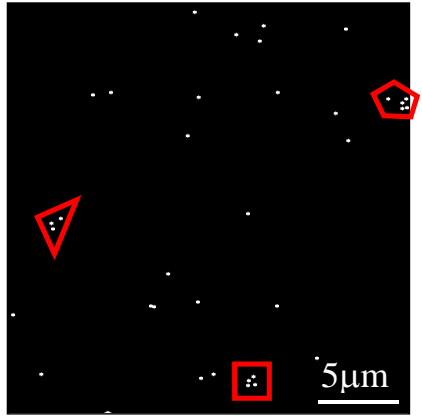

Fig. 3. Confocal image showing three-particle (marked by a triangle), four-particle (marked by a square) and a five-particle cluster (marked by a pentagon), in sample S1. The image is recorded at a depth of 150 μm from the cover-slip using a 40x/0.75 objective.

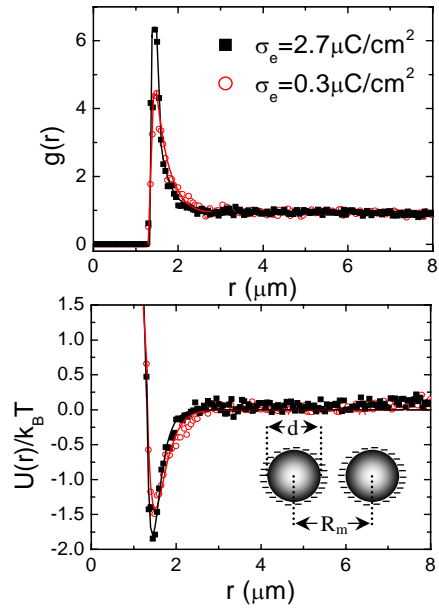

Fig. 4. Pair correlation function $g(r)$ and the corresponding effective pair-potential $U(r)$ (in units of $k_BT$) determined using Eq. (1) for sample S1 (Black) and sample S2 (red). The effective charge density on the particles of sample S1 is one order of magnitude higher than that on the particles of sample S2. The pair-potential $U(r)$ clearly shows an attractive minimum. The minimum (marked in the inset) occurs at $R_m$ ~1.45 μm for sample S1 and ~1.48 μm for sample S2. The well depth is $U_m$ ~ 1.82 $k_BT$ for sample S1 and ~1.46 $k_BT$ for sample S2. In the top panel the lines drawn are guide to the eye. The expression for $U(r)$ (Eq. 2) is fitted to the data and the fits are shown as continuous lines in the bottom panel.